\documentclass[aps,prl,superscriptaddress,twocolumn,showpacs]{revtex4}
\usepackage{amsmath,amsfonts,bm,graphicx,amssymb}
\newcommand{\lso}{\ell_{\mathrm{so}}}
\newcommand{\lo}{\ell_{o}}
\newcommand{\la}{\langle}
\newcommand{\ra}{\rangle}

\begin{document}

\title{Spin-Orbit Mediated Control of Spin Qubits}
\author{Christian Flindt}
\affiliation{Niels Bohr Institute, Universitetsparken 5, DK-2100 Copenhagen, Denmark} \affiliation{MIC --
Department of Micro and Nanotechnology, NanoDTU, Technical University of Denmark, Building 345east, DK-2800
Kongens Lyngby, Denmark}

\author{Anders S. S\o rensen}
\affiliation{Niels Bohr Institute, Universitetsparken 5, DK-2100
Copenhagen, Denmark}

\author{Karsten Flensberg}
\affiliation{Niels Bohr Institute, Universitetsparken 5, DK-2100
Copenhagen, Denmark}
\date{\today}

\begin{abstract}
We propose to use the spin-orbit interaction as a means to control
electron spins in quantum dots, enabling both single qubit and two
qubit operations. Very fast single qubit operations may be
achieved by temporarily displacing the electrons. For two qubit
operations the coupling mechanism is based on a combination of the
spin-orbit coupling and the mutual long-ranged Coulomb
interaction. Compared to existing schemes using the exchange
coupling, the spin-orbit induced coupling is less sensitive to
random electrical fluctuations in the electrodes defining the
quantum dots.
\end{abstract}


\maketitle

It is believed that solid state systems could facilitate
large-scale quantum computing~\cite{NewJPhys:2004} due to the
well-developed fabrication techniques that allow for a high degree
of scalability. On the other hand, solid state systems are
inherently more noisy than, \emph{e.g.}, quantum optical systems,
and in particular several sources of low-frequency noise are
typically present in a solid state environment. One prominent
candidate for solid state quantum computing uses electron spins in
semiconductor quantum dots as carriers of the fundamental unit of
information, the qubit~\cite{Loss:1998}. Electron spins have the
advantage that they are weakly coupled to the surroundings and
therefore weakly sensitive to noise. At the same time, however,
this weak coupling makes the electron spin hard to control
experimentally. To couple two spin qubits, it was proposed to use
the exchange coupling between electron spins in neighboring
quantum dots~\cite{Burkard:1999}, and this was recently
demonstrated experimentally \cite{Petta:2005}. Here the triplet
and singlet spin states have different charge profiles, thereby
enabling electrical control of the coupling. Unfortunately, this
spin-charge coupling also makes the qubits sensitive to electrical
noise and in particular to low-frequency noise \cite{Hu:2006}. In
this setting the spin-orbit interaction is also considered as a
source of decoherence~\cite{Khaetskii:2000}, because it mixes spin
and charge. Recently, however, it has been proposed that it could
also play a role in the coherent interaction of qubits
\cite{Stepanenko:2003}. In this Letter we take these ideas further
and propose to use the spin-orbit interaction as a general means
to manipulate electron spins. The spin-orbit interaction allows
for electrical control of both single and two qubit operations,
but unlike the exchange interaction, the spin-orbit interaction
generates dressed states of spin and charge where the mixing
happens at a high frequency, making the interaction less
susceptible to low-frequency noise.

\begin{figure}
\includegraphics[width=0.44\textwidth, trim = 0 0 0 0, clip]{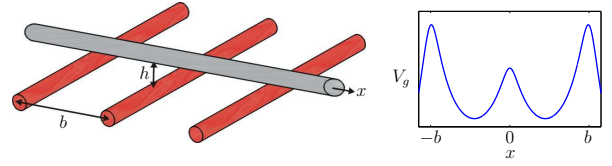}
\caption{(Color online) A nanowire (light gray) placed above three electrodes. The electrodes are used to
define electrostatically a double quantum dot within the nanowire. The electrodes are placed at a distance $b$
apart, while the nanowire is situated at a distance $h$ above the plane of the electrodes. The inset shows a
representative curve for the potential $V_g(x)$ along the nanowire (the $x$-axis) with the electrodes placed
at positions $x=-b$, 0, $b$, respectively. The shown setup resembles the one used in the experiment in Ref.\
\cite{Fasth:2005}. }\label{fig:setup}
\end{figure}

While the general methods we propose in this Letter are applicable in a wide range of situations, we only
consider a simplified one-dimensional model where the  electrons are localized in quantum dots by an external
potential $V(x)$. Physically, we may think of a structure like the one shown schematically in
Fig.~\ref{fig:setup} which was recently realized experimentally \cite{Fasth:2005}, and we give realistic
parameters corresponding to such a system. In our model we include a perpendicular magnetic field $B$
(defining the $z$ direction) and a spin-orbit coupling of the form $p\sigma^y$, where $p$ denotes the momentum
in the $x$ direction \footnote{By a suitable alignment of the magnetic field and definition of the spin
operators this form of the coupling applies to any type of spin-orbit interaction linear in $p$ and may have
contributions from both bulk and structure inversion asymmetry \cite{Levitov:2003}.}. With two electrons
trapped in a double dot potential the Hamiltonian of the system reads
\begin{equation}
\begin{split}
H&=H_1+H_2+\frac{e^2}{4\pi\varepsilon_r\varepsilon_0|x_2-x_1|},\\
H_i&=\frac{p_i^2}{2m}+V(x_i)+\frac{1}{2}g\mu_BB\sigma^z_i+\alpha
p_i\sigma_i^y,\quad i=1,2.
\end{split}
\label{eq:hamiltonian}
\end{equation}
Here $\alpha$ denotes the strength of the spin-orbit coupling,
while $m$ is the effective electron mass. Below, we first consider
how the spin-orbit interaction allows us to control the spin state
of a single electron and then move on to discuss how the
combination of the spin-orbit interaction and the Coulomb
interaction enables two qubit operations in a manner analogous to
the method used for trapped ions~\cite{cirac:2000}.

First, we consider a single electron, and for our analytical
calculation we assume that the potential is harmonic, but has a
time varying equilibrium position denoted $\bar{x}(t)$,
$V(x)=m\omega_0^2(x-\bar{x}(t))^2/2$. We have omitted the
subscript $i$, since we are considering a single electron.
Physically the time varying equilibrium position can be induced
with time varying potentials on the electrodes. We proceed by
performing a unitary transformation $H\rightarrow UHU^{\dagger}$
with $U=\exp{(i \sigma^y(x-\bar x(0))/\lso)}$, where we have
introduced the spin-orbit length $\lso=\hbar/m\alpha$, which
characterizes the length scale of the spin-orbit interaction,
\emph{i.e.}, in the absence of a magnetic field, a spin along the
$x$ or $z$ directions is flipped after traveling a distance
$\pi\lso/2$. With this transformation the Hamiltonian becomes
\begin{equation}
\begin{split}
H&=\frac{p^2}{2m}+\frac{1}{2}m\omega_0^2(x-\bar{x}(t))^2+\frac{1}{2}
g \mu_B B \times \\ &{\left[\cos{\left(\frac{2(x-\bar x(0))}{\lso}\right)}
\sigma^z-\sin{\left(\frac{2(x-\bar x(0))}{\lso}\right)}\sigma^x\right]}.
\end{split}
\label{eq:Hamone}
\end{equation}
We further assume that the renormalized Zeeman splitting
$\Delta_z\equiv\tilde{g}\mu_B B$ ($\tilde{g}$ defined below) is
much smaller than the oscillator energy $\hbar \omega_0$, and that
the equilibrium position is changed adiabatically with respect to
the oscillator frequency $\omega_0\gg (1/\lo)(d \bar x(t)/dt)$,
where $\lo=\sqrt{\hbar/m\omega_0}$ is the characteristic
oscillator length. In this limit, we can trace out the motional
degrees of freedom and obtain
\begin{equation}
\begin{split}
H_{\mathrm{spin}}=\frac{1}{2}\tilde{g}\mu_B B\bigg
[&\cos{\left(\frac{2(\bar{x}(t)-\bar x(0))}{\lso }\right)}\sigma^z
\\ &-\sin{\left(\frac{2(\bar{x}(t)-\bar
x(0))}{\lso}\right)}\sigma^x\bigg],
\end{split}
\label{eq:onedressed}
\end{equation}
with the renormalized $g$-factor \cite{Debald:2005B} given by
\begin{equation}
\tilde{g}=g\left\langle
e^{2i(x-\bar{x}(0))/\lso}\right\rangle=g\exp\!\left[-(\lo/\lso)^2\right].
\label{eq:gdressed}
\end{equation}

The renormalization of the $g$-factor reflects that the qubits states are not pure states of the electron
spin, but dressed states of spin and position \cite{Wu:2003}. Normally, such admixture of spin and position
introduces decoherence because the position is coupled to charge fluctuations, but in this case the charge
distribution is independent of the spin state, and produces no decoherence in the absence of a magnetic field.
With a magnetic field the dressed states are still insensitive to slowly varying electric fields (slow
compared to the Zeeman frequency $\Delta_z/\hbar$), which only shift the equilibrium position, while the spin
state follows adiabatically. The dressed states will, however, be sensitive to slowly varying gradients of the
electric fields which change the trap frequency $\omega_0$. Because of the dependence of $\tilde{g}$ on $\lo$
in Eq.\ \eqref{eq:gdressed} such fluctuations in $\omega_0$ will affect the Zeeman splitting and thereby cause
decoherence of the spin states. As we shall see below this is one of the major limitations for the coupling of
two qubits, but it will not significantly affect the fast single qubit operations, provided that the ground
state width is reasonably well-defined.

Remarkably, the coupling of spin and position can be used to perform single qubit operations if one applies an
electric field with a sufficiently fast variation in time. If we consider the Hamiltonian in Eq.\
\eqref{eq:onedressed} there are two distinct principles for such single qubit operations. One was considered
in Ref.\ \cite{Rashba:2003}, where small amplitude oscillations of the equilibrium position ($|\delta \bar
x(t)|\ll \lso$) at the Zeeman frequency produced spin-orbit induced electron spin resonance (ESR) oscillations
between the two spin states. The second possibility, similar to Ramsey spectroscopy, for producing spin
reversals is to first perform a large rapid change of the equilibrium position for a very short duration. In
particular, if we change $\bar x (t)$ by $\pi \lso/4$ on a time scale much shorter than $\hbar/\Delta_z$, the
effective Hamiltonian becomes $H_{\mathrm{spin}}= \Delta_z \sigma^x/2$. Consequently, after a time
$\hbar\pi/\Delta_z$ the spin state has been flipped and we can then change $\bar x(t)$ back to the original
position. The second procedure has the advantage that it does not rely on any resonance conditions and allows
for very rapid manipulation of the spin. Since the time scale can be changed by changing the magnetic field,
the spin flip time will in practice only be limited by how fast one can change the voltage on the electrodes.
As a particular example of material parameters, we take parameters typical for InAs, \emph{i.e.}, $m=0.027
m_e$, $\hbar\alpha =3\cdot 10^{-8}$ meVm and $g=14.8$, giving $\lso=94$ nm. With $B=40$ mT and
$\hbar\omega_0=0.5$ meV, we get $\lo=75$ nm, so that the Zeeman frequency is $\Delta_z/\hbar=(2\pi) 4.4$ GHz.
The required displacement of the electron spin can be achieved by applying an electric field $E=\pi\lso
m\omega_0^2/4e=6.6 \mu$V/nm, which for a setup like in Fig.\ \ref{fig:setup} with an electrode spacing of
$b=500$ nm corresponds to a voltage of roughly 3 mV, and the entire spin flip process can be achieved in
approximately 0.1 ns.

In addition to being important for performing logical operations
in a quantum computer, the ability to perform rapid spin flips
also allows the reduction of the leading kinds of decoherence due
to the presence of magnetic impurities and the hyperfine coupling
to nuclear spins. For electron spins in GaAs a dephasing time on
the order of 10 ns has been reported \cite{Petta:2005}, and we
expect a similar time scale for InAs. This dephasing can, however,
be reversed by applying pulses, which flip the spin on a much
shorter time scale \cite{Petta:2005}.

The spin-orbit interaction can also mediate \textit{two-qubit interactions} in a very effective way. Returning
to the setup shown in Fig.\ \ref{fig:setup} described by the Hamiltonian in Eq.\ \eqref{eq:hamiltonian}, we
consider next the possibility of using the dipole moment associated with displacements of the electron charges
to couple the two spins. In this one dimensional geometry the two dots holding each one spin are separated by
a distance $d=\bar{x}_2-\bar{x}_1>0$. We consider the limit where the two electrons are well-separated,
allowing us to expand the Coulomb interaction term as $1/|x_2-x_1|\simeq 1/d-\delta/d^2+\delta^2/d^3$, while
using $d\gg\delta\equiv (x_2-\bar{x}_{2})-(x_1-\bar{x}_{1})$. The first term in this expansion gives a
constant contribution to the energy, the second term corresponds to constant forces, which redefine  the two
equilibrium positions. The last term has diagonal terms, $x_i^2$, which provide a small renormalization of the
trapping frequencies. Finally, the interesting term is the cross-term
$-2(x_1-\bar{x}_{1})(x_2-\bar{x}_{2})/d^3$, which results in a coupling of the two orbitals degrees of
freedom, and hence also, via the spin-orbit interaction, between the two spin degrees of freedom.

To calculate this coupling we go back to Eq.\ \eqref{eq:Hamone} (for each electron) with time independent
equilibrium positions $\bar x_i(t)=\bar x_i(0)$, and perform perturbation theory in the magnetic field. To
second order in $B$ the effective Hamiltonian for a single electron spin is still given by Eq.
(\ref{eq:onedressed}), \emph{i.e.}, $H_i=\tilde{g}\mu_BB \sigma^z_i/2$ (for $i=1,2$) plus a spin independent
contribution. The cross-term that couples the two spins gives rise to an effective spin coupling term given by
\begin{equation}\label{Hxx}
    H_{\mathrm{spin},12}=-\frac{e^2}{2\pi\varepsilon_0\varepsilon_rd^3}\langle x_1-\bar{x}_{1}\rangle
    \langle x_2-\bar{x}_{2}\rangle,
\end{equation}
where the brackets only refer to a trace over the vibrational
state, \textit{not} the spin state. Because the two orbital
degrees to this order are decoupled the expectation value
separates. To leading order in $B$, the displacement of the
electrons can be found from Eq.\ \eqref{eq:Hamone} by first order
perturbation theory, and we obtain
\begin{equation}
\la x_i-\bar x_i\ra=\sigma^x_i\frac{\tilde{g}\mu_B B
\lo^2}{\hbar\omega_0\lso}. \label{eq:displacement}
\end{equation}
Combining this with the single particle contributions, we arrive at
the final effective Hamiltonian for the two spins
\begin{equation}
H_{\mathrm{spin}}=\tau\sigma^{x}_1\sigma^{x}_2+\frac{1}{2}\tilde{g}\mu_BB
\left(\sigma^{z}_1+\sigma^{z}_2\right)
\label{eq:spin_hamiltonian},
\end{equation}
where the coupling constant $\tau$ is given by
\begin{equation}
\tau=-\frac{e^2}{4\pi\varepsilon_0\varepsilon_r}\frac{2\lo^4(\tilde{g}\mu_BB)^2}{\lso^2(\hbar\omega_0)^2d^3}.
\label{eq:coupling}
\end{equation}
We stress that this effective Hamiltonian is correct to all orders
in the spin-orbit coupling, but only to second order in the
$B$-field and first order in the Coulomb interaction between the
two electrons. The last approximation can, however, be relaxed
\textit{without} changing the form of the Hamiltonian, but at the
expense of a more complicated expression for $\tau$.

To characterize the stability of the proposed coupling mechanism
to slowly varying perturbations, such as fluctuations in the gate
electrodes, we develop a more realistic model for the double-dot
potential $V(x)$. We are having in mind an experimental setup like
the one shown in Fig.~\ref{fig:setup} \cite{Fasth:2005}, and
consequently we consider the electrostatic potential $V_g(x)$
created by three parallel electrodes with spacing $b$, each
modeled as an infinite line charge, along the $x$-axis running
perpendicular to the electrodes at a distance $h$ above the plane
of the electrodes. The ratio of the charge density on the left
(central) electrode $\lambda_{l(c)}$ and the right electrode
$\lambda_r$ is denoted $\beta_{l(c)}$, i.e.,
$\beta_{l(c)}\equiv\lambda_{l(c)}/\lambda_r$, which we assume can
be controlled via the voltages applied to the electrodes.
Moreover, we define $\hbar\omega_g\equiv e\lambda_r/
4\pi\varepsilon_r\varepsilon_0$ and $x_g\equiv
\sqrt{\hbar/m\omega_g}$, in terms of which we express the
electrostatic potential as
\begin{equation}
V_g(x)=V_e(x+b,\beta_l)+V_e(x,\beta_c)+V_e(x-b,1) \label{eq:gatepotential}
\end{equation}
with $V_e(x,\beta)=-\beta\hbar\omega_g\ln{\left[(x^2+h^2)/x_g^2\right]}$. A representative curve for $V_g(x)$
is shown the inset of Fig.~\ref{fig:setup}.

\begin{figure}
\includegraphics[height=0.185\textwidth, trim = 0 0 0 0, clip]{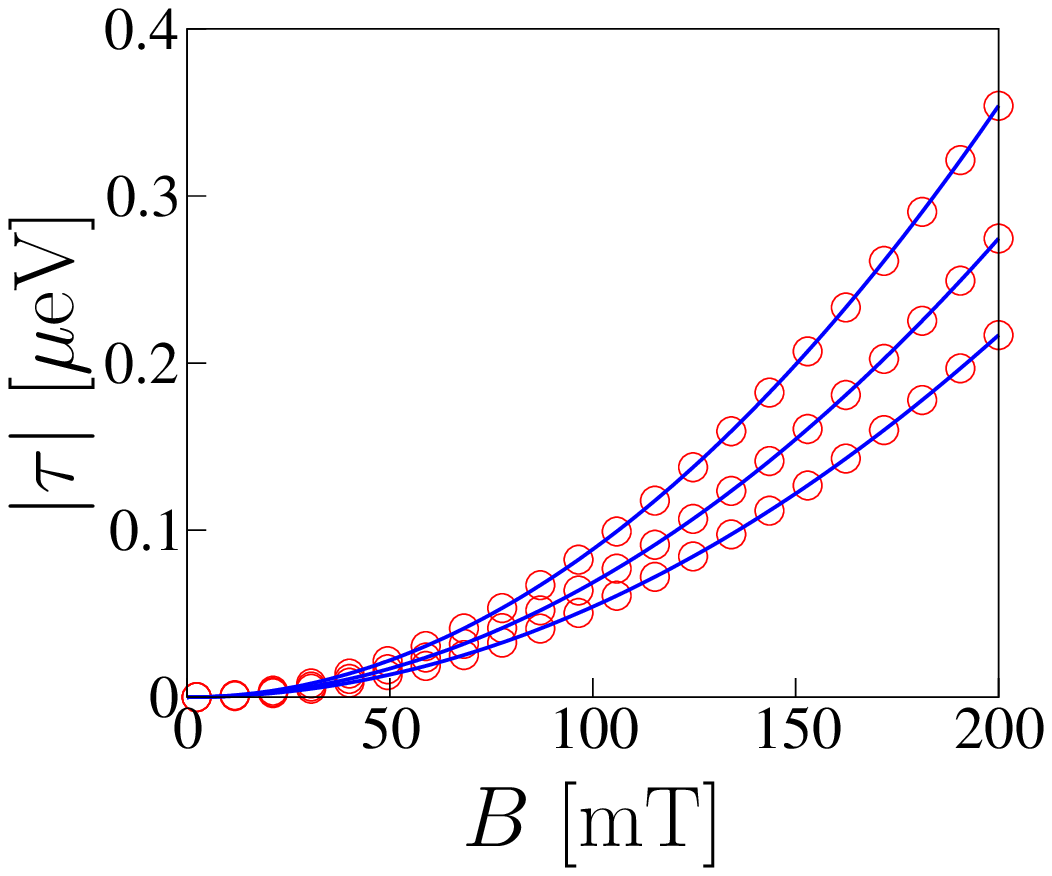}
\includegraphics[height=0.18\textwidth, trim = 0 0 0 0, clip]{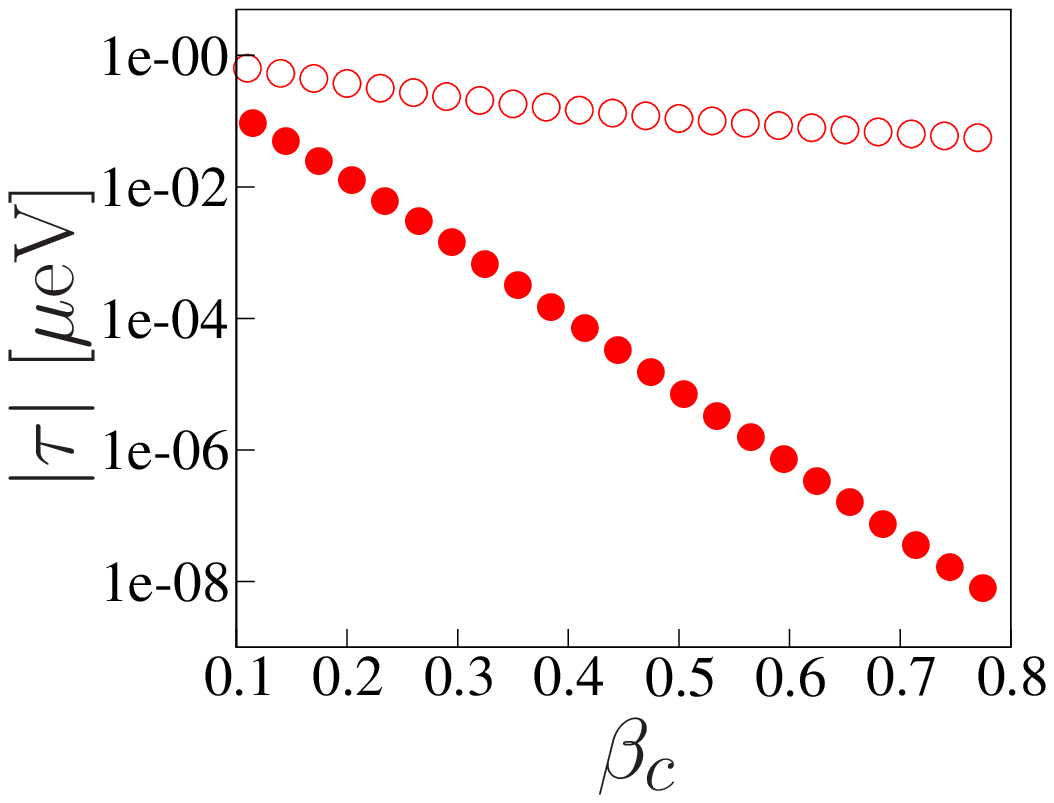}
\caption{(Color online) Numerical calculation of the coupling~$\tau$. Material parameters correspond to InAs
and we use $\hbar\omega_g=1$~meV, $\beta_l=1$, $x_g=53$~nm, $h=0.5x_g$, $b=10x_g$. Left panel: The coupling
$\tau$ as function of applied magnetic field $B$ for $\beta_c=0.6$ (upper circles), 0.7, 0.8 (lower circles).
Solid lines show Eq.\ \eqref{eq:coupling} with the orbital energy spacing $\hbar\omega_0$ extracted from the
low-energy spectrum and $d$ being the only fitting-parameter. Corresponding to $\beta_c=0.6, 0.7, 0.8$, we
have $\hbar\omega_0=0.39,0.40,0.42$~meV, and $d=8.3x_g, 8.7x_g, 9.1x_g$, respectively. Right panel: Open
circles show the coupling $\tau$ as a function of the applied voltage on the central electrode parametrized by
$\beta_c$ with $B=40$ mT. Solid circles show the contribution from the bare exchange coupling $J$.}
\label{fig:numerics1}
\end{figure}

We have implemented numerically on a finite-size real-space grid the two-particle Hamiltonian in Eq.\
\eqref{eq:hamiltonian} using $V_g(x)$ in Eq.~\eqref{eq:gatepotential} as the potential $V(x)$. With
$N\sim~100-500$ grid-points, the resulting matrix representation of the Hamiltonian is large (dimension
$4N^2\times 4N^2$), but sparse, allowing for computationally cheap calculations of the low-energy spectrum
from which we can extract the coupling of the various spin states. In the left panel of
Fig.~\ref{fig:numerics1} we show numerical calculations of the coupling $\tau$ as a function of applied
$B$-field. The renormalized Zeeman splitting due to the applied $B$-field is much smaller than the spacing of
the orbital levels, and we thus expect Eq.~\eqref{eq:coupling} to hold. The numerical results show excellent
agreement with the analytic expression. For the parameters used in the figure typical interaction strengths
are $\tau/\hbar=(2\pi)f$, with $f\sim$ 0.1~GHz, corresponding to gate times on the order of $1/(2f)\sim$ 5 ns.

In the right panel of Fig.~\ref{fig:numerics1} we show numerical results for the coupling $\tau$ as a function
of the voltage applied to the central electrode parametrized by $\beta_c$. In order to determine the
contribution arising from the bare exchange coupling (due to the Pauli principle and the Coulomb interaction),
we also show numerical results for the splitting of the spin states without the spin-orbit coupling. Compared
to the bare exchange coupling $J$, which is clearly exponentially dependent on the applied voltage, the
spin-orbit induced coupling shows a weaker voltage dependence.

\begin{figure}
\includegraphics[height=0.19\textwidth, trim = 0 0 0 0, clip]{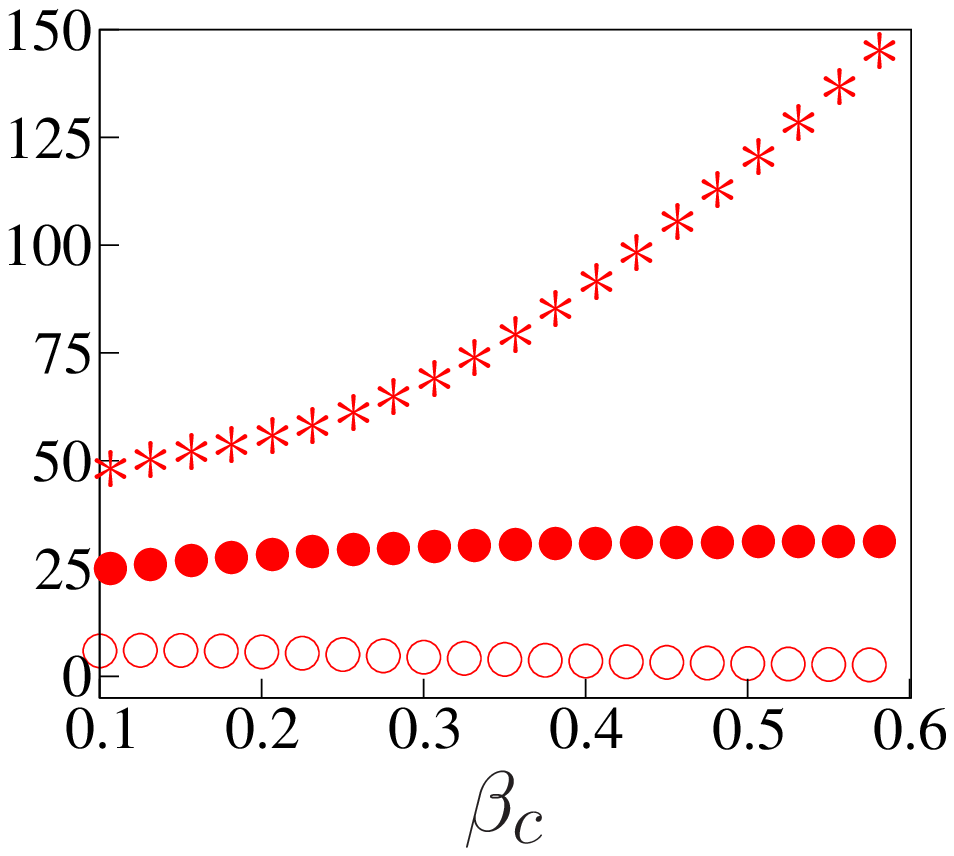}
\includegraphics[height=0.19\textwidth, trim = 0 0 0 0,clip]{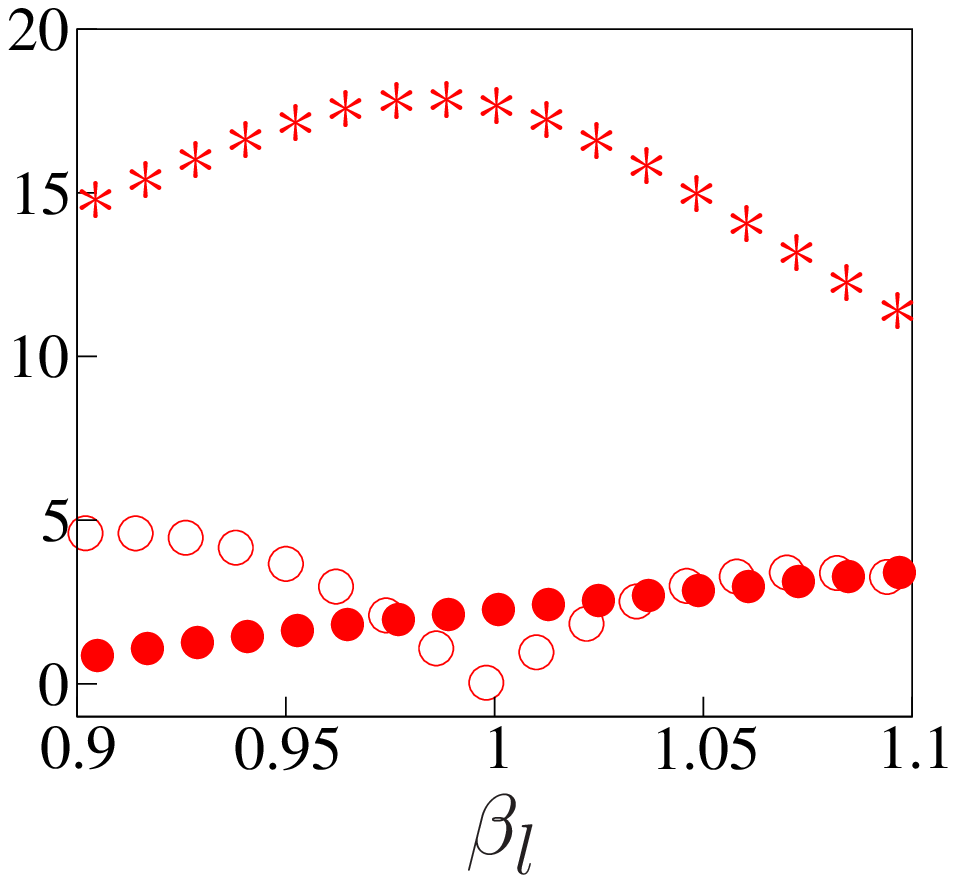}
\caption{(Color online) Numerical calculations of the sensitivity to fluctuations in the gate electrodes. The
sensitivity of the spin-orbit induced coupling $\tau$ and the Zeeman splitting $\Delta_z$ are quantified by
$|(1/\tau)(\partial \tau/\partial \beta_{i})|, i=c,l$ (open circles) and $|(1/\tau)(\partial \Delta_z/\partial
\beta_{i})|, i=c,l$ (stars), respectively. Material parameters correspond to InAs. The other parameters are
$B=80$~mT, $\hbar\omega_g=1$~meV, $x_g=53$~nm, $h=0.5x_g$, and $b=10x_g$. In both panels solid circles show
the sensitivity of the bare exchange coupling $|(1/J)(\partial J/\partial \beta_{i})|, i=c,l$. Left panel:
Sensitivity to fluctuations in the central electrode. The electrostatic potential is symmetric, \emph{i.e.},
$\beta_l=1$. Right panel: Sensitivity to fluctuations in the left electrode with $\beta_c=0.7$.}
\label{fig:numerics2}
\end{figure}

Fluctuations in the electrostatic environment cause fluctuations of the orbital level splitting
$\hbar\omega_0$ and the distance $d$. Typically these fluctuations have the form of $1/f$-noise and
concentrating on the dominating low-frequency component, we characterize in the following the sensitivity of
the coupling to electrical fluctuations using a purely static calculation by taking derivatives with respect
to $\beta_c$ and $\beta_l$. For the spin-orbit induced coupling $\tau$ given in Eq.\ \eqref{eq:coupling}, we
have $|(1/\tau)(\partial\tau/\partial\omega_0)\delta\omega_0|=4|\delta\omega_0/\omega_0|$ and
$|(1/\tau)(\partial\tau/\partial d)\delta d|=3|\delta d/d|$. In order to perform reliable two qubit
operations, both of these quantities must be much smaller than unity, which for the fluctuations imply
$|\delta \omega_0/\omega_0|,|\delta d/d| \ll 0.1$. In Fig.\ \ref{fig:numerics2}, we show $|(1/\tau)(\partial
\tau/\partial \beta_{i})|$ as a function of $\beta_i,i=c,l$. The coupling is stable for $|(1/\tau)(\partial
\tau/\partial \beta_{i})\delta \beta_i|\ll 1,i=c,l$, implying $|\delta\beta_i|< 0.1, i=l,c$ according to the
numerical results. This does not impose any unrealistic requirements on the experimental setup. For comparison
we also show $|(1/J)(\partial J/\partial \beta_{i})|,i=c,l$  for the exchange interaction $J$, which for
fluctuations in $\beta_c$ is an order of magnitude more sensitive.

As discussed previously electrical fluctuations also cause
fluctuations of the renormalized Zeeman splitting, $\Delta_z$. For
$\Delta_z$, we have $|(1/\tau)(\partial
\Delta_z/\partial\omega_0)\delta\omega_0|=(4\pi\varepsilon_r\varepsilon_0\hbar
md^3\omega_0 ^3/e^2\tilde{g}\mu_BB)|\delta\omega_0/\omega_0|$. For
InAs with $d=500$ nm, $\hbar\omega_0=0.3$ meV, $B=80$ mT,
$|(1/\tau)(\partial
\Delta_z/\partial\omega_0)\delta\omega_0|\simeq
50|\delta\omega_0/\omega_0|$, implying the stricter condition
$|\delta\omega_0/\omega_0|\ll 0.01$. In Fig.\ \ref{fig:numerics2},
we show numerical results for $|(1/\tau)(\partial
\Delta_z/\partial \beta_{i})|$ as a function of $\beta_i,i=c,l$.
The results indicate that we must require $|\delta\beta_i|< 0.01,
i=l,c$ for the renormalized Zeeman splitting to be stable. If this
requirement cannot be met in experiments, the problem may be
circumvented by encoding a single qubit in a singlet-triplet pair
as recently discussed in Refs.\ \cite{Petta:2005,Taylor:2005} or
alternatively by combining the gate with fast spin-echo pulses
implemented by rapidly shifting the position of the electrons.

In conclusion, we have presented a spin-orbit induced mechanism
for coherent control of spin qubits in quantum dots. The
spin-orbit coupling allows for fast single qubit operations, and
the two qubit operations are robust against electrical
fluctuations in the electrodes defining the double dot. We
emphasize that although some of the above conclusions have been
made in connection with a specific experimental setup in mind,
they also hold at a more general level.

The authors thank  G.\ Burkard, X.\ Cartoixa, W.\ A.\ Coish, A.\
Fuhrer, A.-P.\ Jauho, M.\ D.\ Lukin, J.\ Nyg\aa rd, and J.\ M.\
Taylor for valuable discussions and comments. This work was
supported by the Danish Natural Science research council.


\begin{thebibliography}{18}
\expandafter\ifx\csname
natexlab\endcsname\relax\def\natexlab#1{#1}\fi
\expandafter\ifx\csname bibnamefont\endcsname\relax
  \def\bibnamefont#1{#1}\fi
\expandafter\ifx\csname bibfnamefont\endcsname\relax
  \def\bibfnamefont#1{#1}\fi
\expandafter\ifx\csname citenamefont\endcsname\relax
  \def\citenamefont#1{#1}\fi
\expandafter\ifx\csname url\endcsname\relax
  \def\url#1{\texttt{#1}}\fi
\expandafter\ifx\csname
urlprefix\endcsname\relax\def\urlprefix{URL }\fi
\providecommand{\bibinfo}[2]{#2}
\providecommand{\eprint}[2][]{\url{#2}}

\bibitem[{New(2005)}]{NewJPhys:2004}
\bibinfo{journal}{For a recent special issue on solid-state quantum computing,
  we refer to New J. Phys., Vol.} \textbf{\bibinfo{volume}{7}}
  (\bibinfo{year}{2005}).

\bibitem[{\citenamefont{Loss and DiVincenzo}(1998)}]{Loss:1998}
\bibinfo{author}{\bibfnamefont{D.}~\bibnamefont{Loss}} \bibnamefont{and}
  \bibinfo{author}{\bibfnamefont{D.~P.} \bibnamefont{DiVincenzo}},
  \bibinfo{journal}{Phys. Rev. A} \textbf{\bibinfo{volume}{57}},
  \bibinfo{pages}{120 } (\bibinfo{year}{1998}).

\bibitem[{\citenamefont{Burkard et~al.}(1999)\citenamefont{Burkard, Loss, and
  DiVincenzo}}]{Burkard:1999}
\bibinfo{author}{\bibfnamefont{G.}~\bibnamefont{Burkard}},
  \bibinfo{author}{\bibfnamefont{D.}~\bibnamefont{Loss}}, \bibnamefont{and}
  \bibinfo{author}{\bibfnamefont{D.~P.} \bibnamefont{DiVincenzo}},
  \bibinfo{journal}{Phys. Rev. B} \textbf{\bibinfo{volume}{59}},
  \bibinfo{pages}{2070 } (\bibinfo{year}{1999}).

\bibitem[{\citenamefont{Petta et~al.}(2005)\citenamefont{Petta, Johnson,
  Taylor, Laird, Yacoby, Lukin, Marcus, Hanson, and Gossard}}]{Petta:2005}
\bibinfo{author}{\bibfnamefont{J.~R.} \bibnamefont{Petta \emph{et al.}}},
\bibinfo{journal}{Science}
  \textbf{\bibinfo{volume}{309}}, \bibinfo{pages}{2180 }
  (\bibinfo{year}{2005}).

\bibitem[{\citenamefont{Hu and Das~Sarma}(2006)}]{Hu:2006}
\bibinfo{author}{\bibfnamefont{X.}~\bibnamefont{Hu}} \bibnamefont{and}
  \bibinfo{author}{\bibfnamefont{S.}~\bibnamefont{Das~Sarma}},
  \bibinfo{journal}{Phys. Rev. Lett.} \textbf{\bibinfo{volume}{96}},
  \bibinfo{pages}{100501} (\bibinfo{year}{2006}).

\bibitem[{\citenamefont{Khaetskii and Nazarov}(2000)}]{Khaetskii:2000}
\bibinfo{author}{\bibfnamefont{A.~V.} \bibnamefont{Khaetskii}}
  \bibnamefont{and} \bibinfo{author}{\bibfnamefont{Yu.~V.}
  \bibnamefont{Nazarov}}, \bibinfo{journal}{Phys. Rev. B}
  \textbf{\bibinfo{volume}{61}}, \bibinfo{pages}{12639 }
  (\bibinfo{year}{2000}),
\bibinfo{author}{\bibfnamefont{A.~V.} \bibnamefont{Khaetskii}}
  \bibnamefont{and} \bibinfo{author}{\bibfnamefont{Yu.~V.}
  \bibnamefont{Nazarov}}, \bibinfo{journal}{Phys. Rev. B}
  \textbf{\bibinfo{volume}{64}}, \bibinfo{pages}{125316}
  (\bibinfo{year}{2001}).

\bibitem[{\citenamefont{Stepanenko et~al.}(2003)\citenamefont{Stepanenko,
  Bonesteel, DiVincenzo, Burkard, and Loss}}]{Stepanenko:2003}
\bibinfo{author}{\bibfnamefont{D.}~\bibnamefont{Stepanenko \emph{et al.}}},
  \bibinfo{journal}{Phys. Rev. B} \textbf{\bibinfo{volume}{68}}, \bibinfo{pages}{115306}
  (\bibinfo{year}{2003}),
\bibinfo{author}{\bibfnamefont{D.}~\bibnamefont{Stepanenko}} \bibnamefont{and}
  \bibinfo{author}{\bibfnamefont{N.~E.} \bibnamefont{Bonesteel}},
  \bibinfo{journal}{Phys. Rev. Lett.} \textbf{\bibinfo{volume}{93}}, \bibinfo{pages}{140501}
  (\bibinfo{year}{2004}),
\bibinfo{author}{\bibfnamefont{L.-A.}~\bibnamefont{Wu}} \bibnamefont{and}
  \bibinfo{author}{\bibfnamefont{D.~A.} \bibnamefont{Lidar}},
  \bibinfo{journal}{Phys. Rev. A} \textbf{\bibinfo{volume}{66}}, \bibinfo{pages}{062314}
  (\bibinfo{year}{2002}),
\bibinfo{author}{\bibfnamefont{S.}~\bibnamefont{Debald}} \bibnamefont{and}
  \bibinfo{author}{\bibfnamefont{C.}~\bibnamefont{Emary}},
  \bibinfo{journal}{Phys. Rev. Lett.} \textbf{\bibinfo{volume}{94}}, \bibinfo{pages}{226803}
  (\bibinfo{year}{2005}).

\bibitem[{\citenamefont{Fasth et~al.}(2005)\citenamefont{Fasth, Fuhrer,
  Bj{\"{o}}rk, and Samuelson}}]{Fasth:2005}
\bibinfo{author}{\bibfnamefont{C.}~\bibnamefont{Fasth}},
  \bibinfo{author}{\bibfnamefont{A.}~\bibnamefont{Fuhrer}},
  \bibinfo{author}{\bibfnamefont{M.~T.} \bibnamefont{Bj{\"{o}}rk}},
  \bibnamefont{and}
  \bibinfo{author}{\bibfnamefont{L.}~\bibnamefont{Samuelson}},
  \bibinfo{journal}{Nano Lett.} \textbf{\bibinfo{volume}{5}},
  \bibinfo{pages}{1487 } (\bibinfo{year}{2005}).

\bibitem[{\citenamefont{Cirac and Zoller}(2000)}]{cirac:2000}
\bibinfo{author}{\bibfnamefont{J.~I.} \bibnamefont{Cirac}} \bibnamefont{and}
  \bibinfo{author}{\bibfnamefont{P.}~\bibnamefont{Zoller}},
  \bibinfo{journal}{Nature} \textbf{\bibinfo{volume}{404}}, \bibinfo{pages}{579
  } (\bibinfo{year}{2000}).

\bibitem[{\citenamefont{Debald and Kramer}(2005)}]{Debald:2005B}
\bibinfo{author}{\bibfnamefont{S.}~\bibnamefont{Debald}} \bibnamefont{and}
  \bibinfo{author}{\bibfnamefont{B.}~\bibnamefont{Kramer}},
  \bibinfo{journal}{Phys. Rev. B} \textbf{\bibinfo{volume}{71}},
  \bibinfo{pages}{115322} (\bibinfo{year}{2005}).

\bibitem[{\citenamefont{Wu and Lidar}(2003)}]{Wu:2003}
\bibinfo{author}{\bibfnamefont{L.-A.}~\bibnamefont{Wu}} \bibnamefont{and}
  \bibinfo{author}{\bibfnamefont{D.~A.}~\bibnamefont{Lidar}},
  \bibinfo{journal}{Phys. Rev. Lett.} \textbf{\bibinfo{volume}{91}},
  \bibinfo{pages}{097904} (\bibinfo{year}{2003}).

\bibitem[{\citenamefont{Rashba and Efros}(2003)}]{Rashba:2003}
\bibinfo{author}{\bibfnamefont{E.~I.} \bibnamefont{Rashba}} \bibnamefont{and}
  \bibinfo{author}{\bibfnamefont{A.~L.} \bibnamefont{Efros}},
  \bibinfo{journal}{Phys. Rev. Lett.} \textbf{\bibinfo{volume}{91}},
  \bibinfo{pages}{126405} (\bibinfo{year}{2003}),
\bibinfo{author}{\bibfnamefont{V.~N.} \bibnamefont{Golovach}},
  \bibinfo{author}{\bibfnamefont{M.}~\bibnamefont{Borhani}}, \bibnamefont{and}
  \bibinfo{author}{\bibfnamefont{D.}~\bibnamefont{Loss}}
  (\bibinfo{year}{2006}), \eprint{cond-mat/0601674}.

\bibitem[{\citenamefont{Taylor et~al.}(2005)\citenamefont{Taylor, Engel,
  D{\"u}r, Yacoby, Marcus, Zoller, and Lukin}}]{Taylor:2005}
\bibinfo{author}{\bibfnamefont{J.~M.} \bibnamefont{Taylor \emph{et al.}}},
  \bibinfo{journal}{Nature Physics} \textbf{\bibinfo{volume}{1}},
  \bibinfo{pages}{177} (\bibinfo{year}{2005}).

\bibitem[{\citenamefont{Levitov and Rashba}(2003)}]{Levitov:2003}
\bibinfo{author}{\bibfnamefont{L.~S.} \bibnamefont{Levitov}} \bibnamefont{and}
  \bibinfo{author}{\bibfnamefont{E.~I.} \bibnamefont{Rashba}},
  \bibinfo{journal}{Phys. Rev. B} \textbf{\bibinfo{volume}{67}},
  \bibinfo{pages}{115324} (\bibinfo{year}{2003}).
\end{thebibliography}

\end{document}